\documentstyle [epsfig] {article}

\begin {document}

\title{NEW ORIGIN FOR SPIN CURRENT AND CURRENT-INDUCED SPIN PRECESSION IN MAGNETIC MULTILAYERS}

\author{L. BERGER}

\date{Physics Department, Carnegie Mellon University, Pittsburgh, PA 15213}

\maketitle

\setlength{\baselineskip}{4mm}

In metallic ferromagnets, an electric current is accompanied by a flux of angular momentum, also called spin current. In multilayers, spatial variations of the spin current correspond to drive torques exerted on a magnetic layer. These torques result in spin precession above a certain current threshold. The usual kind of spin current is associated with translation of the spin-up and spin-down Fermi surfaces in momentum space. We discuss a different kind of spin current, associated with expansion and contraction of the Fermi surfaces. It is more nonlocal in nature, and may exist even in locations where the electrical current density is zero. It is larger than the usual spin current, in a ratio of 10 or 100, at least in the case of one-dimensional current flow. The new spin current is proportional to the difference $\Delta\overline{\mu}\simeq 10^{-3} eV$ between spin-up and spin-down Fermi levels, averaged over the entire Fermi surface. Conduction processes, spin relaxation, and spin-wave emission in the multilayer can be described by an equivalent electrical circuit resembling an unbalanced dc Wheatstone bridge. And $\Delta\overline{\mu}$ corresponds to the output voltage of the bridge.

\hspace{6em} I. INTRODUCTION

\vspace{1ex}

In magnetic multilayers, the s-d exchange interaction causes $^{1}$ an increase of the Gilbert spin-damping parameter $\alpha$ near an interface. In the presence of a dc current normal to layers, it also generates $^{1-5}$ drive torques on magnetic spins near the interface; these lead to a spin precession above a certain current threshold. This current-induced precession has been observed recently $^{6-8}$ in Co/Cu multilayers. Such device, which we call a SWASER, is the magnetic analog of the semiconductor laser diode. Current-induced switching between opposite static magnetic-spin directions has also been observed $^{7,8}$, at low magnetic fields; it happens probably through a temporary current-induced precession. Microwave-frequency ac currents have also been used $^{9}$ to excite precession in a multilayer. Detailed numerical as well as analytical calculations of dc current-induced precession and switching have been done $^{10}$ for various thin-film geometries and anisotropy-field cases. A physically different mechanism has been proposed $^{11}$ for current-induced switching.  

The layer configuration most common in theory and experiments is shown in Fig. 1a. Magnetic layer $F_{1}$ prepares the spin of conduction electrons in a fixed specific direction. It must be $^{1}$ thicker than a spin-diffusion length, in order to ensure a ratio $\alpha_{1}$ of spin-up to spin-down current as different from one as possible. The non-magnetic spacer N must be thinner than a local spin-diffusion length, to preserve spin memory for electrons crossing it. The precessing magnetic layer $F_{2}$ must be thin, as the current-induced drive torque is only applied $^{2}$ to a region of $F_{2}$ of thickness $\pi/(k_{\uparrow}-k_{\downarrow})$ near the $N-F_{2}$ interface. Here, $k_{\uparrow}$ and $k_{\downarrow}$ are the spin-up and spin-down Fermi wavenumbers in $F_{2}$, both assumed real-valued. Thick non-magnetic layer $N_{2}$ acts as a current-return lead. We introduce $^{1}$ spatial coordinates x,y,z, with x normal to layers, and x=0 at the $N-F_{2}$ interface (Fig. 1a). We also use spin coordinates $x_{2},y_{2},z_{2}$, with $z_{2}$ parallel to the precessing atomic spins $\vec{S}_{2}$ in $F_{2}$, and $x_{2}$ in the plane of $z_{2}$ and of the precession axis (Fig. 1b). The direction of the precession axis, which is also the direction of the total effective magnetic field, is assumed parallel to the fixed atomic spins $\vec{S}_{1}$ in $F_{1}$. The $\vec{S}_{1}$ direction is arbitrary. The precession-cone angle between $\vec{S}_{1}$ and $\vec{S}_{2}$ is called $\theta$.

The current-induced torque $\vec{\tau}$ on the spins of layer $F_{2}$, responsible for maintaining the precession, can be calculated $^{4}$ from the angular-momentum current $\vec{P}$, also called spin current. The important components are along $x_{2}$:

\begin{equation}
\tau_{x2}=P_{x2}(x=0)-P_{x2}(x=\infty).
\end{equation}

Here, $P_{x2}(x=0)$ is calculated in $F_{2}$ at the $N-F_{2}$ interface, and $P_{x2}(x=\infty)$ at a point of $F_{2}$ to the right of the region $0<x<\pi/(k_{\uparrow}-k_{\downarrow})$ where most of the torque is applied.

The traditional spin current $^{4}$ arises from Fermi-surface translation, and is proportional to the local electrical-current density. The main purpose of the present paper is to describe a different kind of spin current, which comes from Fermi-surface expansion or contraction, and which can exist even where the current density is zero; this new spin current turns out to be much larger than the traditional one, at least when the current flow is one-dimensional. Another purpose is to demonstrate the equivalence of the Slonczewski and SWASER theories.

\vspace{1ex}

\hspace{3em} II. SPIN CURRENT FOR ONE ELECTRON

\vspace{1ex}

First, we consider free-electron states of wavevector $\vec{k}_{N}$ in N which are eigenstates of the $z_{2}$ component of spin. If $k^{N}_{x}>0$, they enter $F_{2}$ from N, and their form in $F_{2}$ is

\begin{eqnarray}
\psi_{\uparrow}=\left | \begin{array} {c} 1 \\ 0 \end{array} \right | f_{\uparrow}(\vec{r}); \ \psi_{\downarrow}=\left | \begin{array} {c} 0 \\ 1 \end{array} \right | f_{\downarrow}(\vec{r}) \\
f_{\uparrow}= \frac{exp(i\vec{k}_{\uparrow}\vec{r})}{V_{N}^{1/2}}\frac{2k^{N}_{x}}{k^{N}_{x}+k^{\uparrow}_{x}}; \ f_{\downarrow}= \frac{exp(i\vec{k}_{\downarrow}\vec{r})}{V_{N}^{1/2}}\frac{2k^{N}_{x}}{k^{N}_{x}+k^{\downarrow}_{x}}. \nonumber
\end{eqnarray}

The wavevectors $\vec{k}_{\uparrow},\vec{k}_{\downarrow}$ are such that $\psi_{\uparrow},\psi_{\downarrow}$ have the same energy. These states are assumed to have most of their norm in the layer N of volume $V_{N}$. Momentum conservation along y and z at the $N-F_{2}$ interface gives $k^{N}_{y}=k^{\uparrow}_{y}=k^{\downarrow}_{y}$ and $k^{N}_{z}=k^{\uparrow}_{z}=k^{\downarrow}_{z}$. We neglect reflections at the $F_{2}-N_{2}$ interface.

A part of the amplitude of the actual spin-up and spin-down waves $\psi_{+},\psi_{-}$ entering $F_{2}$ from N with $k^{N}_{x}>0$ originates $^{1}$ in $F_{1}$. Hence, the spin density $<\vec{s}\delta(\vec{r}-\vec{r}_{0})>$ for that part is parallel or antiparallel to $\vec{S}_{1}$, and at an angle $\theta$ to the $z_{2}$ axis. These states are linear combinations of $\psi_{\uparrow}, \psi_{\downarrow}$. With $z_{2}$ as a spin quantization axis as before, we can write in $F_{2}$:

\begin{equation}
\psi_{+}=T^{1/2}_{+}\left| \begin{array} {c} f_{\uparrow}(\vec{r})cos(\theta/2) \\ -f_{\downarrow}(\vec{r})sin(\theta/2)\end{array} \right | ;  \psi_{-}=T^{1/2}_{-}\left| \begin{array} {c} f_{\uparrow} (\vec{r})sin(\theta/2) \\ f_{\downarrow}(\vec{r})cos(\theta/2) \end{array} \right |.
\end{equation}

We assume $\theta \ll 1$ radian, so that $\psi_{+},\psi_{-}$ can still  be called spin-up and spin-down. The quantities $T_{+}=4k^{\uparrow}_{x}k^{N}_{x}/(k^{N}_{x}+k^{\uparrow}_{x})^{2}, T_{-}=4k^{\downarrow}_{x}k^{N}_{x}/(k^{N}_{x}+k^{\downarrow}_{x})^{2}$ are the transmission coefficients at the $F_{1}-N$ interface, assuming $F_{1}$ to be made of the same material as $F_{2}$. These quantities are present because we consider only the parts of the waves in $F_{2}$ originating in $F_{1}$. To simplify the theory, we approximate $T_{+}\simeq T_{-}\simeq T$.

For any state $\psi$, Slonczewski gives $^{4}$ an expression for the components $P_{x2},P_{y2}$ of the spin current:

\begin{equation}
\psi=\left | \begin{array} {c} a(\vec{r}) \\ b(\vec{r}) \end{array} \right |;\  P_{x2}+iP_{y2}=\frac{i\hbar^{2}}{2m}(\frac{da^{*}}{dx}b-a^{*}\frac{db}{dx}).
\end{equation}

Using this on the $k^{N}_{x}>0$ states $\psi_{+},\psi_{-}$ of Eqs.(3), we obtain in $F_{2}$, respectively:

\begin{eqnarray}
(P_{x2}(x))_{+} & = & -\frac{\hbar}{4}sin\theta\frac{T(v^{\uparrow}_{x}+v^{\downarrow}_{x})(2k^{N}_{x})^{2}cos(k^{\uparrow}_{x}-k^{\downarrow}_{x})x}{V_{N}(k^{N}_{x}+k^{\uparrow}_{x})(k^{N}_{x}+k^{\downarrow}_{x})} \\
 (P_{x2}(x))_{-} & = & -(P_{x2}(x))_{+}. \nonumber
\end{eqnarray}
 
Here, $v^{\uparrow}_{x}$ and $v^{\downarrow}_{x}$ are electron-velocity components in $F_{2}$. The cosine factor comes from the precession $^{1,4}$ of the electron spin around the s-d exchange field set up by $\vec{S}_{2}$, as the electron moves toward increasing x in $F_{2}$. We assume $k_{N}<k_{\uparrow},k_{\downarrow}$ so that the $k^{N}_{x}>0$ states are not totally reflected at the $N-F_{2}$ interface.

On the other hand, the $k^{N}_{x}<0$ states originated in $N_{2}$ before traversing $F_{2}$ toward left. Because of the absence of a magnetic layer like $F_{1}$ to polarize their spin along a specific fixed direction oblique to $z_{2}$, we have $P_{x2}=0$ for these states, at any location in $F_{2}$. The same applies to the part of the states in $F_{2}$ that have $k^{N}_{x}>0$ after reflection at the $N-F_{2}$ interface. Also to the part (ignored earlier) that has $k^{N}_{x}>0$ after reflection at the $F_{1}-N$ interface, if $T_{+}\simeq T_{-}$.

In thermal equilibrium, $k^{N}_{x}>0$ spin-up and spin-down states $\psi_{+},\psi_{-}$ of the same energy are equally populated, and Eqs. (5) show that their $P_{x2}$ cancel each other in pairs. The total $P_{x2}$ of the electron distribution vanishes, as expected from the second law since $P_{x2}$ is related to damping and drive torques. This cancellation does not happen in general for component $P_{z2}$, which plays no important role anyway.

For the same reason, when out of equilibrium, only states near the Fermi level contribute to the total $P_{x2}$. 

\vspace{1ex}

\hspace{3em} III. TOTAL SPIN CURRENT IN CASE OF

\hspace{5em}  TRANSLATED FERMI SURFACES

\vspace{1ex}

The traditional calculation $^{4}$ of the total spin current assumes a non-zero local current density $j_{x}$, corresponding to translations of the spin-up and spin-down Fermi surfaces in N along the $k^{N}_{x}$ axis, by amounts $\delta k^{+}_{x}, \delta k^{-}_{x}$ (Fig. 2a).

Because of the cancellation mentioned above, only states located between the spin-up and spin-down Fermi surfaces in N contribute to the total $P_{x2}$, at zero temperature. Also, the blurring of the electron distribution near the Fermi level, happening at non-zero temperature, turns out to have no effect on the total $P_{x2}$.

We integrate the $P_{x2}$ of Eqs. (5) over all number elements $d^{2}N$ of electron states, from $\alpha=0$ to $\alpha=\pi/2$, corresponding to the $k^{N}_{x}>0$ half of the Fermi surface in N. For given energy, $k^{\uparrow}_{x},k^{\downarrow}_{x}$ are functions of $\alpha$. As explained in the last section, only the states on that half have non-zero $P_{x2}$. We call $\Sigma P_{x2}$ the total spin current:

\begin{equation}
\Sigma P_{x2}(x)=-\frac{\hbar^{2} sin\theta}{16\pi^{2}m} k^{2}_{N}(\delta k^{+}_{x}-\delta k^{-}_{x})\int^{\pi/2}_{0}T\frac{(2k^{N}_{x})^{2}(k^{\uparrow}_{x}+k^{\downarrow}_{x})cos((k^{\uparrow}_{x}-k^{\downarrow}_{x})x)cos\alpha sin\alpha d\alpha}{(k^{N}_{x}+k^{\uparrow}_{x})(k^{N}_{x}+k^{\downarrow}_{x})}.
\end{equation}

For any given $x\gg \pi/(k_{\uparrow}-k_{\downarrow})$, the argument of $cos((k^{\uparrow}_{x}-k^{\downarrow}_{x})x)$ in the integral varies rapidly with $\alpha$. Hence, the average of this cosine over half a Fermi surface is nearly zero. In turn, this means $\Sigma P_{x2}(x)\simeq0$ when $x\gg \pi/(k_{\uparrow}-k_{\downarrow})$. Thus, the last term of Eq. (1) vanishes. Only $\Sigma P_{x2}(x=0)$ matters.

An alternate, but equivalent, approach $^{1}$ to the problem of current-induced torques replaces the spin current by the spin imbalance, defined as the difference $\Delta \mu=\mu_{\uparrow}-\mu_{\downarrow}\simeq 10^{-4}\ eV$ between the spin-up and spin-down Fermi levels at a given point of the Fermi surface. As before, only the electrons on the $k^{N}_{x}>0$ half matter, because some of them have their spin along $\vec{S}_{1}$, and because they enter $F_{2}$. 

When $\Delta\mu<0$, the quantity $\Delta\mu$ ``pumps'' the electron gas, just like the difference between valence-band and conduction-band Fermi levels does $^{12}$  in a laser diode, resulting in stimulated emission of spin waves, i.e., in precession. The current threshold where precession begins is given $^{1,6}$ by $\Delta\mu+\hbar\omega=0$, where $\omega$ is the angular frequency of precession, at any temperature. A mathematically identical condition applies $^{12}$ to the threshold of a laser diode.

In the case of Fermi-surface translation, $\Delta\mu$ is proportional to   $\delta k^{+}_{x}-\delta k^{-}_{x}$ or to $j_{x}$:

\begin{equation}
\Delta\mu=-2\frac{\alpha_{1}-1}{\alpha_{1}+1}\frac{\hbar k^{N}_{x}}{en^{N}_{e}}\frac{T_{av}}{2}j_{x}=\frac{\hbar^{2}}{m}k^{N}_{x}(\delta k^{+}_{x}-\delta k^{-}_{x})\frac{T_{av}}{2}.
\end{equation}

Here, $\alpha_{1}=j^{\uparrow}_{x}/j^{\downarrow}_{x}$ is the local ratio of spin-up to spin-down current densities in N, and $n^{N}_{e}$ the electron density per unit volume in N. We average $k^{N}_{x}$ over the $k^{N}_{x}>0$ half of the Fermi surface, obtaining $k^{N}_{x}\simeq k_{N}/2$. Similarly, $T_{av}$ is some average of T over that half. As before, the $T_{av}/2$ factor is present because only transmitted $k^{N}_{x}>0$ waves are active for SWASER action. We combine Eqs. (6,7) and obtain

\begin{equation}
\frac{\Sigma P_{x2}(x=0)}{\Delta\mu} = -\frac{k_{N}sin\theta}{8\pi^{2}}\frac{2}{T_{av}}\int^{\pi/2}_{0}T\frac{(2k^{N}_{x})^{2}(k^{\uparrow}_{x}+k^{\downarrow}_{x})\ cos\alpha\ sin\alpha\ d\alpha}{(k^{N}_{x}+k^{\uparrow}_{x})(k^{N}_{x}+k^{\downarrow}_{x})}.
\end{equation}

We see that $\Sigma P_{x2}$ and $\Delta\mu$ vary in a constant ratio when $j_{x}$ or $\alpha_{1}=j^{\uparrow}_{x}/j^{\downarrow}_{x}$ changes. This reflects the fact that the Slonczewski theory, based on $\Sigma P_{x2}$, is nearly equivalent to the SWASER theory based on $\Delta\mu$. The only significant difference is that the first theory uses only bulk damping, and the second theory only surface damping. 

\vspace{1ex}

\hspace{3em}  IV. TOTAL SPIN CURRENT IN CASE OF EXPANDED

\hspace{5em}   OR CONTRACTED FERMI SURFACE

\vspace{1ex}

There exists a different kind of spin imbalance, associated with isotropic expansion or contraction of the Fermi surface rather than translation (Fig 2b). It is the one originally introduced by Aronov $^{13}$ and by Johnson and Silsbee $^{14}$. This quantity, which we denote $^{3}$ by $\Delta\overline{\mu}$, is an average of $\mu_{\uparrow}-\mu_{\downarrow}$ over the whole Fermi surface, not just one half. Since it depends on $j^{\uparrow}_{x}$ and $j^{\downarrow}_{x}$ values over a range of x of the order of a spin-diffusion length, we call it the nonlocal spin imbalance. It may exist even at locations where the electrical current density is zero. Like $\Delta\mu<0$, $\Delta\overline{\mu}<0$ can be used $^{3,6}$ to induce spin precession in a multilayer. For the configuration of Fig. 1a, $\Delta\overline{\mu}$ has the same sign as $\Delta\mu$.

We can calculate $^{3}$ $\Delta\overline{\mu}$ for the multilayer of Fig. 1a at the $N-F_{2}$ interface by solving the spin-diffusion equations in one dimension, assuming the $F_{1}$ and $N_{2}$ thicknesses to be much larger than a local spin-diffusion length $l_{sr}$, and the $N,F_{2}$ thicknesses much smaller than a local $l_{sr}$. A formula for $\Delta\overline{\mu}$ was given as Eq. (5) of Ref. 3. By comparing the formula to our Eq. (7) for $\Delta\mu$, one obtains roughly

\begin{equation}
\frac{\Delta\mu}{\Delta\overline{\mu}}\leq(\frac{\Lambda_{\downarrow}}{l_{sr}})_{F1}+(\frac{\Lambda_{\downarrow}}{l_{sr}})_{N2}.
\end{equation}

Here, $\Lambda_{\downarrow}$ and $l_{sr}$ are the spin-down mean free path and the spin-diffusion length, and F1, N2 refer to layers $F_{1}$ and $N_{2}$. We see from Eq. (9) that the $\Delta\overline{\mu}$ mechanism is usually dominant over the $\Delta\mu$ mechanism, since we have $\Lambda_{\downarrow}/l_{sr}\simeq \ 0.1-0.01$ in most magnetic materials $^{15}$. This is helped by the fact that $\Lambda_{\downarrow}<\Lambda_{\uparrow}$ in Co, Ni, Ni-Fe.

The spin-current component $P_{x2}$ of one electron is still given by Eq. (5). It is still true that only electrons located between the spin-up and spin-down Fermi surfaces (Fig. 2b) contribute to the total spin current of the electron distribution, which we denote now by $\Sigma\overline{P}_{x2}$ to differentiate it from the translation case. We consider a momentum-space volume element $d^{2}k$ at an angle $\alpha$ to the $k^{N}_{x}$ axis, containing $d^{2}N$ states of one spin:

\begin{eqnarray}
d^{2}k=k^{2}_{N}\ 2\pi\ sin\alpha\ d\alpha\ \delta k_{N} \\
d^{2}N=V_{N}d^{2}k/(8\pi^{3}). \nonumber
\end{eqnarray}

Here, $\delta k_{N}$ is the radial distance between Fermi surfaces, now independent of $\alpha$. It is related to $\Delta\overline{\mu}$ by

\begin{equation}
\Delta\overline{\mu}=\frac{\hbar^{2}}{m}k_{N}\ \delta k_{N}\frac{T^{'}_{av}}{2}.
\end{equation}

The quantity $T^{'}_{av}$ is a slightly different average of T than $T_{av}$.Using Eqs. (10, 11), we integrate the $P_{x2}$ of Eqs. (5) over all number elements $d^{2}N$ to obtain $\Sigma\overline{P}_{x2}$. As before, only the $k^{N}_{x}>0$ half of the Fermi surface contributes:
 
\begin{equation}
\frac{\Sigma\overline{P}_{x2}(x=0)}{\overline{\Delta\mu}(x=0)} = -\frac{k_{N}sin\theta}{16\pi^{2}}\frac{2}{T^{'}_{av}}\int^{\pi/2}_{0}T\frac{(2k^{N}_{x})^{2}(k^{\uparrow}_{x}+k^{\downarrow}_{x})\ sin\alpha\ d\alpha}{(k^{N}_{x}+k^{\uparrow}_{x})(k^{N}_{x}+k^{\downarrow}_{x})}.
\end{equation}

This formula for Fermi-surface expansion or contraction differs from that of Eq. (8) for translation by an extra 1/2 factor in the front, as well as by a missing $cos \alpha$ in the integral. Therefore, the ratios of spin current to spin imbalance given by the two equations are roughly the same. Just as $\Delta\overline{\mu}$ dominates over $\Delta\mu$, $\Sigma\overline{P}_{x2}$ is dominant over $\Sigma P_{x2}$ if one-dimensional flow is assumed. This implies a considerable reduction of the threshold current where precession starts.

As before, we have $\Sigma\overline{P}_{x2}\simeq 0$ for $x \gg \pi/(k_{\uparrow}-k_{\downarrow})$. Only $\Sigma\overline{P}_{x2}(x=0)$ matters in Eq. (1). 

Most multilayer samples used in recent SWASER experiments $^{6-8}$ have current leads flaring out immediately to much wider effective cross section area than that of the multilayer itself. Thus, current flow is not one-dimensional. Because of the increased volume available for spin relaxation in these leads, $\Delta\overline{\mu}$ and $\Sigma\overline{P}_{x2}$ are considerably reduced below their value in one dimension assumed in the present paper, for given current density in layer $F_{2}$. Also, layer $F_{1}$ is too thin, leading to the same result. On the other hand, $\Delta\mu$ and $\Sigma P_{x2}$ in $F_{2}$ are not affected by these factors.

In order to maximize $|\Delta\overline{\mu}|$ and $|\Sigma\overline{P}_{x2}|$  and to verify the predictions of the present paper, the cross-section area of layer $N_{2}$, which plays the role of a lead (Fig. 1a), should remain equal to that of layer $F_{2}$, over distances from layer $F_{2}$ at least equal to a local spin-diffusion length $l_{sr}$. And the $F_{1}$ thickness should be at least a local $l_{sr}$,  again with constant cross section area. Like us, Fert and Lee $^{16}$ find that $\Delta\overline{\mu}$ is not determined by interface scattering in the multilayer itself.

The importance of $\Delta\overline{\mu}$ and $\Sigma\overline{P}_{x2}$ can be compared to that of $\Delta\mu$ and $\Sigma P_{x2}$ by measuring the threshold current of such one-dimensional sample, and comparing to that of a traditional sample with more multidimensional flow. Another way would be to realize the so-called zero-current SWASER $^{3}$, which is based on $\Delta\overline{\mu}$ and $\Sigma\overline{P}_{x2}$ alone.

\vspace{1ex}

\hspace {1em} V. EQUIVALENT ELECTRICAL CIRCUIT

\vspace{1ex}

Electrical conduction in the multilayer of Fig. 1a can be described by an equivalent dc electrical circuit (Fig. 3). It may be used to calculate $\Delta \overline{\mu}$.

The upper horizontal resistors represent conduction in the spin-down band in $F_{1}$ and $N_{2}$, including both interface and bulk scattering, and the lower horizontal resistors conduction in the spin-up band. The vertical resistors represent electron-spin relaxation between the two bands. The dc electrical motor represents the interaction between electrons and spin waves in $F_{2}$. The dc current through the motor corresponds to the rate of electron spin flip in $F_{2}$ associated $^{1}$ with spin-wave generation, i.e., with inducing spin precession. And $\Delta\overline{\mu}/e$ at the $N-F_{2}$ interface is represented by the dc voltage between points D and U. The counter-electromotive force of the motor is equal to $\hbar \omega/e$, and its internal resistance, shown as $R_{M}$, is inversely proportional to the number $n_{m}$ of spin-wave quanta (magnons) in $F_{2}$. This circuit (Fig.3) can be justified on the basis of the spin-diffusion equations $^{3}$ and of the formula $^{1}$ for the rate of production of magnons in $F_{2}$.

Consideration of the equivalent circuit, like that of the spin-diffusion equations $^{3}$, shows that the SWASER is affected by conduction processes in $F_{1}$ and $N_{2}$ over a range of x of at least one spin-diffusion length, in one dimension. It also shows that $\Delta\mu$ and $\Sigma P_{x2}$ could  be maximized, for given $j_{x}$, in the case of Fermi-surface translation by having a very short spin-relaxation time in layer $N_{2}$ as compared to other layers. On the other hand, a long relaxation time in all layers is best in the case of expansion or contraction. This (large) dependence of $\Sigma P_{x2}$ on spin relaxation in the various layers is absent from Eqs. (31-33) of Waintal et al. $^{17}$, since they ignore spin relaxation and the spin-diffusion equations. Also, they ignore $N_{2}$.

\vspace{1ex}

\hspace{1em}   VI. CONCLUSIONS

\vspace{1ex}

In the present paper, we demonstrated the existence of a new kind of spin current, connected with contraction and expansion of the Fermi surface. For one-dimensional current flow, it is larger than the traditional spin current based on translation of the Fermi surface. Each spin current is proportional to its own kind of spin imbalance. Finally, we introduced an equivalent dc electrical circuit describing the processes of electrical conduction, spin relaxation and spin-wave emission in a magnetic multilayer traversed by a dc electrical current perpendicular to layers.
 
I thank Dr Ya. Bazaliy and Dr M. Tsoi for very useful discussions.

\vspace{1ex}

\hspace{5em}   REFERENCES

\vspace{1ex}

1. L. Berger, Phys. Rev. B $\underline{54}$, 9353 (1996).

2. L. Berger, J. Appl. Phys. $\underline{81}$, 4880 (1997).

3. L. Berger, IEEE Trans. Magn. $\underline{34}$, 3837 (1998).

4. J.C. Slonczewski, J. Magn. Magn. Mater. $\underline{159}$, L1 (1996); J. Magn. Magn. Mater. $\underline{195}$, L261 (1999).

5. Ya.B. Bazaliy, B.A. Jones and S.C. Zhang, Phys. Rev. B $\underline{57}$, R3213 (1998). 

6. M. Tsoi, A.G.M. Jansen, J. Bass, W.-C. Chiang, M. Seck, V. Tsoi and P. Wyder, Phys. Rev. Lett. $\underline{80}$, 4281 (1998); $\underline{81}$, 493(E) (1998); M. Tsoi, A.G.M. Jansen, J. Bass, W. C. Chiang, V. Tsoi and P. Wyder, Nature $\underline{406}$, 46 (2000).

7. E.B. Myers, D.C. Ralph, J.A. Katine, R.N. Louie and R.A. Buhrman, Science $\underline{285}$, 867 (1999); J.Z. Sun, J. Magn. Magn. Mater. $\underline{202}$, 157 (1999).

8. J.A. Katine, F.J. Albert, R.A. Buhrman, E.B. Myers and D.C. Ralph, Phys. Rev. Lett. $\underline{84}$, 3149 (2000); F.J. Albert, J.A. Katine, R.A. Buhrman and D.C. Ralph, Appl. Phys. Lett. $\underline{77}$, 3809 (2000).

9. S.M. Rezende, F.M. de Aguiar, M.A. Lucena and A. Azevedo, Phys. Rev. Lett. $\underline{84}$, 4212 (2000).

10. J.Z. Sun, Phys. Rev. B $\underline{62}$, 570 (2000); Ya.B. Bazaliy, B.A. Jones and S.-C. Zhang, preprint cond-mat/0009034.

11. C. Heide, P.E. Zilberman and R.J. Elliott, preprint cond-mat/0005064.
 
12. M.G.A. Bernard and G. Duraffourg, Phys. Stat. Solidi $\underline{1}$, 699 (1961).

13. A.G. Aronov, JETP Letters, $\underline{24}$, 32 (1976).

14. M. Johnson and R.H. Silsbee, Phys. Rev. Lett. $\underline{55}$, 1790 (1985).
 
15. T. Valet and A. Fert, Phys. Rev. B, $\underline{48}$, 7099 (1993); A. Fert, J.L. Duvail and T. Valet, Phys. Rev. B $\underline{52}$, 6513 (1995).

16. A. Fert and S.-F. Lee, Phys. Rev. B, $\underline{53}$, 6554 (1996).

17. X. Waintal, E.B. Myers, P.W. Brouwer and D.C. Ralph, Phys. Rev. B, $\underline{62}$, 12317 (2000).

\vspace{1ex}

\hspace{3em} FIGURE CAPTIONS

\vspace{1ex}

FIG. 1. a) Multilayer used for current-driven eperiments. Thick magnetic layer $F_{1}$ with fixed magnetization prepares the spin of conduction electrons in a specific direction. On the other side of thin non-magnetic spacer N are the thin precessing magnetic layer $F_{2}$ and the thick non-magnetic layer $N_{2}$. b) Definition of spin coordinates $x_{2}, y_{2}, z_{2}$ in relation to the fixed atomic spin $\vec{S}_{1}$ in $F_{1}$ and to the precessing atomic spin $\vec{S}_{2}$ in $F_{2}$.

FIG. 2. a) The circles represent the spin-up and spin-down Fermi surfaces in N. A local electrical current density $j_{x}>0$ causes translations by amounts $\delta k^{+}_{x}<0, \delta k^{-}_{x}<0$ of these Fermi surfaces along $k_{x}$, with $\Delta\mu<0$. b) Now, one Fermi surface is expanded and the other contracted. The radial distance $\delta k_{N}$ of the Fermi surfaces is proportional to the nonlocal spin imbalance $\Delta\overline{\mu}<0$, according to Eq. (11).

FIG. 3. Equivalent dc electrical circuit for conduction processes in the multilayer of Fig. 1a. The upper horizontal resistors represent conduction in the spin-down band in the various layers, and the lower horizontal resistors in the spin-up band. The vertical resistors represent electron-spin relaxation between the two bands. The dc electrical motor represents the interaction between electrons and spin waves in $F_{2}$.

\newpage

\begin{figure}

\epsfig{file=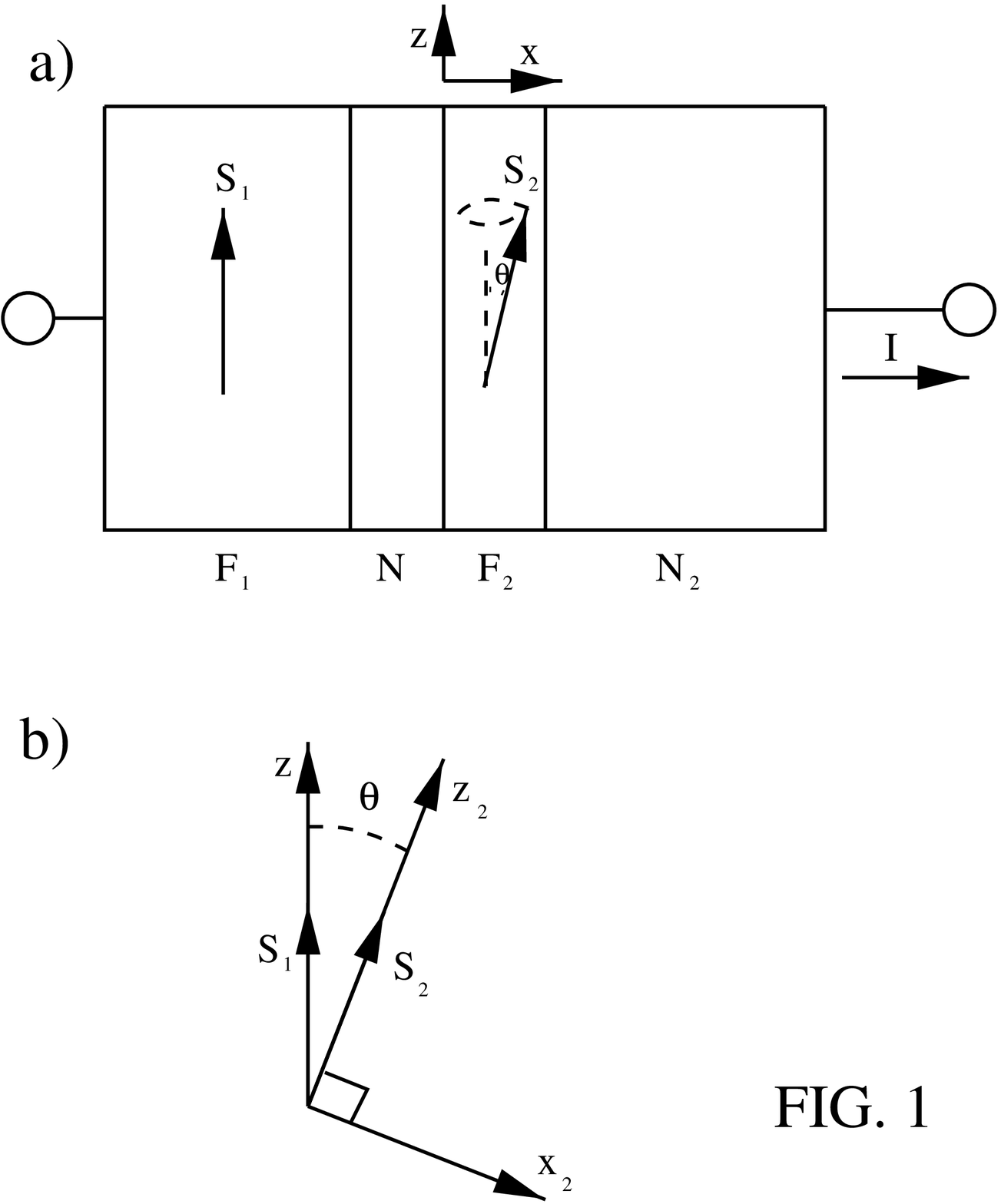,width=4in}

\end{figure}

\vspace{20mm}

\begin{figure}

\epsfig{file=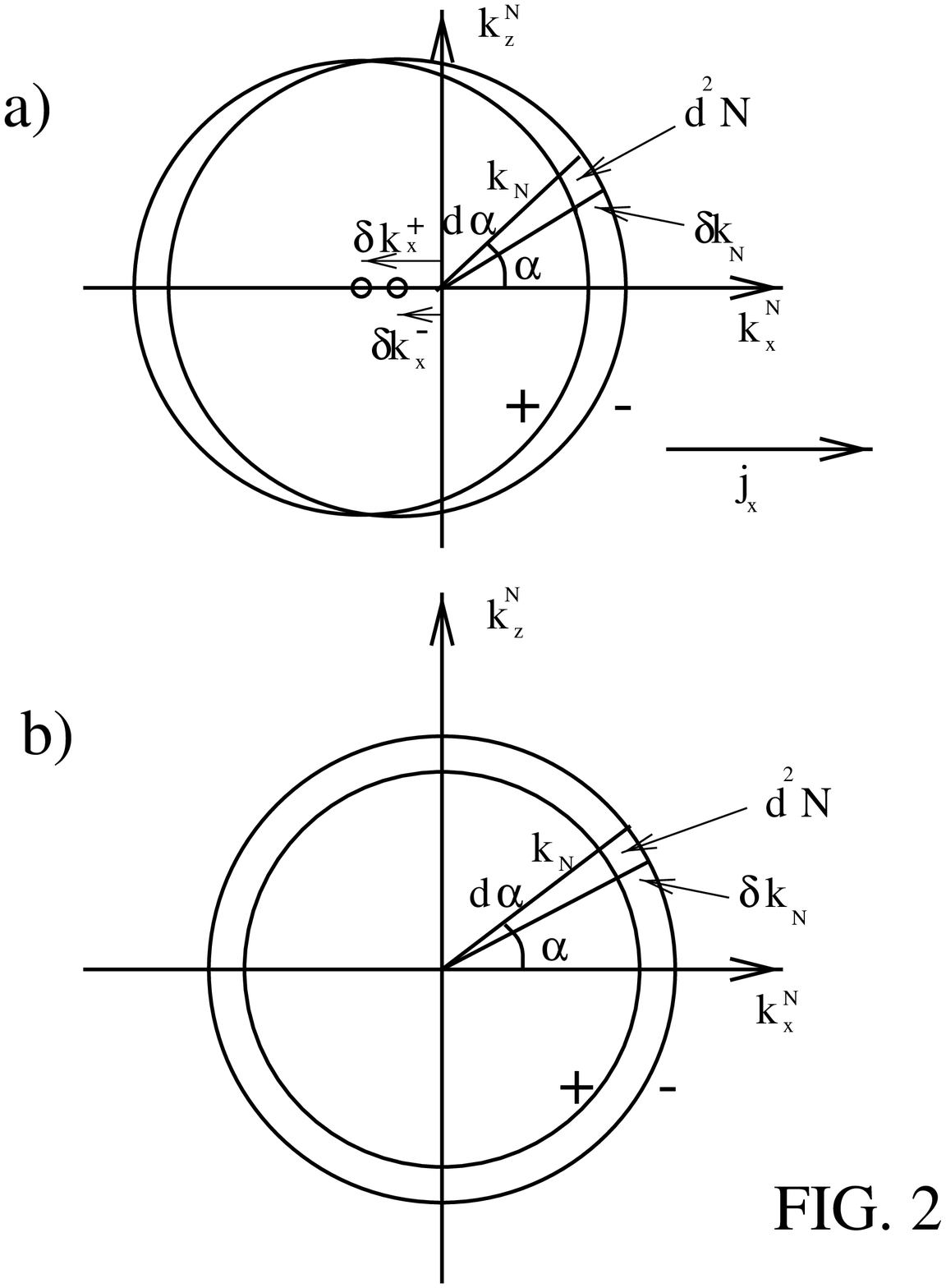,width=4in}

\end{figure}

\vspace{20mm}

\begin{figure}

\epsfig{file=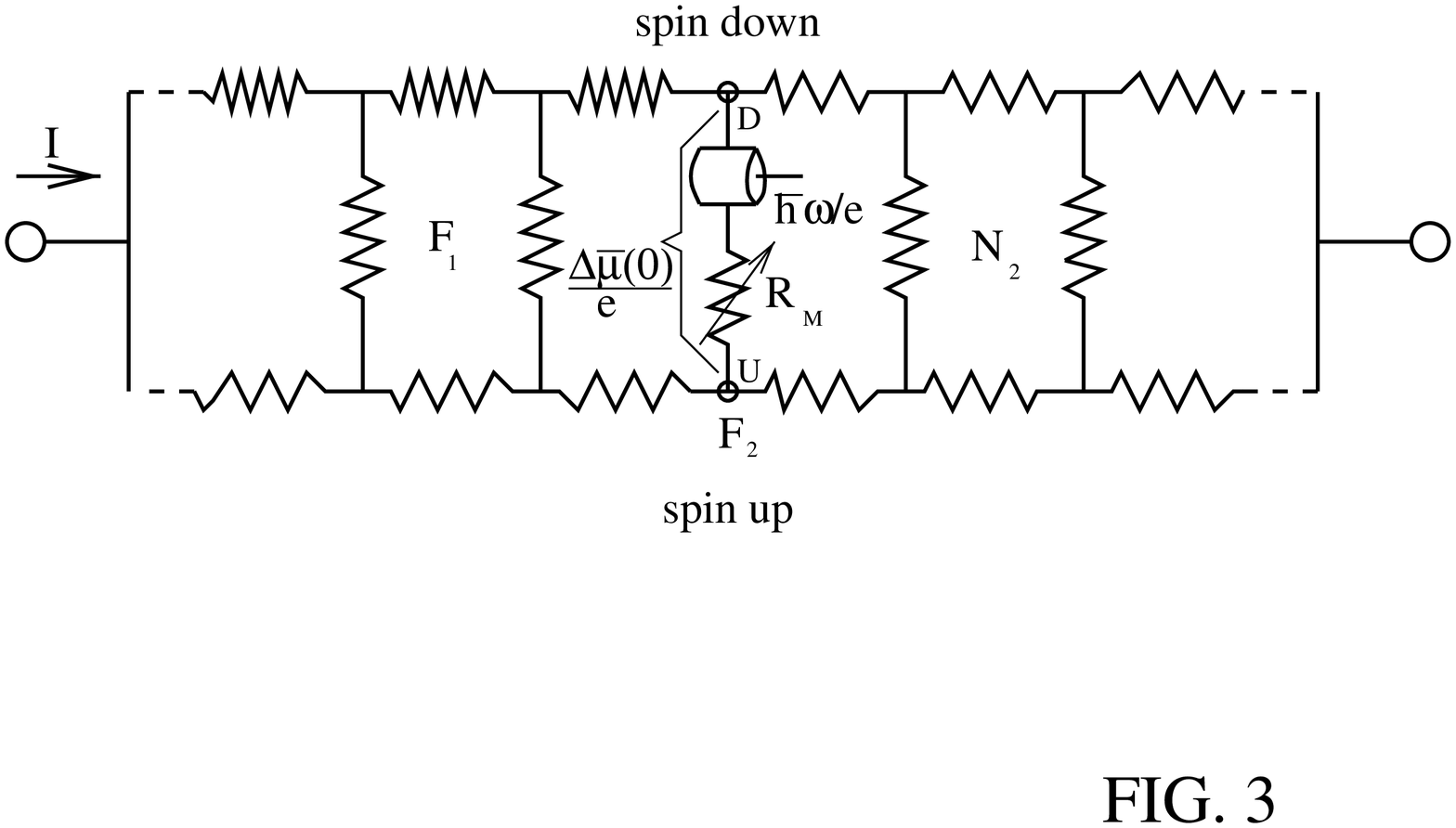,width=4in}

\end{figure}

\end{document}